\documentclass[aps,tightenlines,showpacs]{revtex4}
\usepackage{graphicx}
\usepackage{float}
\usepackage[T1]{fontenc}
\usepackage[latin1]{inputenc}
\usepackage{amssymb}
\include{epsf}
\epsfverbosetrue

\newcommand{\bea}{\begin{eqnarray}}
\newcommand{\eea}{\end{eqnarray}}
\newcommand{\Qsl}{Q\hspace{-.22cm}/~}

\newcommand{\Rsl}{R\hspace{-.22cm}/~}
\newcommand{\Ksl}{K\hspace{-.22cm}/~}
\newcommand{\rhatsl}{\hat r \hspace*{-.17cm}/}
\newcommand{\xx}{\noindent}
\newcommand{\ra}{\rightarrow}

\makeatother
\begin{document}

\title{The soft fermion dispersion relation at next-to-leading order in Hot QED}

\author{M.E. Carrington}

\affiliation{Department of Physics, Brandon University,
Brandon, Manitoba, R7A 6A9 Canada\\
and Winnipeg Institute for Theoretical Physics,
Winnipeg, Manitoba, Canada}

\begin{abstract}
We study next-to-leading order contributions to the soft static fermion dispersion relation in hot QED. 
We derive an expression for the complete next-to-leading order contribution to the retarded fermion self-energy. The real and imaginary parts of this expression give the next-to-leading order contributions to the mass and damping rate of the fermionic quasi-particle. Many of the terms that are expected to contribute according to the traditional power counting argument are actually subleading. We explain why the power counting method over estimates the contribution from these terms. 
For the electron damping rate in QED we obtain: $\gamma_{QED} = \frac{e^2 T}{4\pi}(2.70)$. We check our method by calculating the next-to-leading order contribution to the  damping rate for the case of QCD with two flavours and three coulours. Our result agrees with the result obtained previously in the literature. The numerical evaluation of the nlo contribution to the mass is left to a future publication. 
\end{abstract}


\maketitle

\section{Introduction}

It is well known that the behaviour of elementary particles becomes modfied when the particle propagates in a medium. The particles become ``dressed'' by their interaction with the medium, and one speaks of collective modes, or quasi-particles. 
One studies these collective modes by looking at the corresponding thermal propagators. The behaviour of the quasi-particles is deduced from the analytic structure of the propagator. 
In \cite{KKR} it was shown from general principles that the singularity structure of certain components of gauge and matter propagators are gauge-independent, when all contributions of a given order are systematically taken into account.

However, the actual calculation of dispersion relations for soft quantities at next-to-leading order is notoriously difficult. The original paper by Braaten and Pisarski \cite{BP} identified three potential contributions at next-to-leading order. They are: (1) corrections to the result for the 1-loop diagram obtained by expanding to next-to-leading order in the ratio of the external momentum to the loop momentum; (2) contributions to the 2-loop diagrams from the region of the phase space that corresponds to both loops hard; and (3) the contribution to 1-loop diagrams with soft loop momentum, and all propagators and vertices replaced with htl effective ones. We will refer to these three types of contributions as (1-terms), (2-terms), and (3-terms). We note that the power counting arguments of Braaten and Pisarski refer to the maximum possible contribution from each type of term. The actual contribution may be lower order for kinematical reasons, or because of some cancellation between different integrals. 

Only a few calculations of next-to-leading order quantities have been done. The gluon damping rate was calculated by Braaten and Pisarski \cite{BPgluondamp}. The damping rate of a soft static quark was calculated by Kobes, Kunstatter and Mak \cite{KKM}, and by Braaten and Pisarski \cite{BPquarkdamp}. All three of these calculations involve only the imaginary part of the appropriate 2-point vertex function. The real part is generally more difficult to calculate. The gluonic plasma frequency in the long wavelength limit has been calculated by Schulz \cite{SCH}, and the Debye mass has been studied in \cite{Tonydebye}.  

In this paper we consider the next-to-leading order corrections to the dispersion relation for soft static fermions $Q_\mu=(q_0\sim e T,\vec q=0)$. For static fermions, the collective modes appear mathematically as poles of the propagator. The mass and damping rate of the quasi-particle are obtained from the solution of the equation:
\bea
{\rm det}\;(\Qsl-{\bf \Sigma}_{ret}(Q))\Big|_{q_0=M-i\gamma} = 0\,.
\eea
The leading contributions are obtained from the hard thermal loop (htl) results:
\bea
\label{leadingorder}
m_f = \sqrt{e^2T^2/8}\,;~~\gamma^{(0)} = 0\,.
\eea
The goal is to find an expression for the next-to-leading order corrections to this result.

One of the main results of this paper is a compact expression for the next-to-leading order contributions to the retarded fermion self-energy in QED. This result can be divided into real and imaginary parts (Eqns. (\ref{imag-part}) and (\ref{real-part})) which give, respectively, the next-to-leading order corrections to the mass and damping rate in (\ref{leadingorder}).
The expression for the imaginary part can be checked by converting to the corresponding QCD expression (obtained previously in \cite{KKM,BPquarkdamp}) by calculating the appropriate  group factors.  The expression for the real part has not appeared previously in the literature. 

The imaginary part of the next-to-leading (nlo) self-energy gives the nlo contribution to the damping rate \cite{Blaizot}: 
\bea
\label{damp}
\gamma = -\frac{1}{4}{\rm Tr}(\gamma_0 {\rm Im} {\bf \Sigma}(q_0,0))\Big|_{q_0=m_f} \,.
\eea
To check our method, we have calculated the nlo quark damping rate for QCD with 2 flavours and 3 colours and obtained a result that agrees with that of \cite{KKM,BPquarkdamp}. (Note that these authors use a different convention for the definition of the damping rate: they define the damping rate with an extra factor of 1/2 relative to the definition used here). 
For QCD with 2 flavours and 3 colours we obtain: 
\bea
\{N_f=2\,,~N_c=3\}~\Rightarrow~\gamma = \frac{g^2 T C_F}{4\pi}(2.81)\,;~~C_F = 4/3\,.
\eea
The corresponding result for QED has not been previously calculated. The result is:
\bea
\gamma_{QED} = \frac{e^2 T}{4\pi}(2.70)\,.
\eea

The calculation of the real part of the fermion self-energy is more complicated for several reasons which we discuss below. 

(1) There are pure real tadpole type contributions that can be dropped in the calculation of the imaginary part. 

(2) The leading order contribution to the imaginary part of the self-energy is zero (because of kinematic constraints), and thus the complete evaluation of the integral gives a contribution of $e^2T$ times some number of order one, which is the full nlo contribution. For the real part, the numerical result for the integral will contain the leading order term (which is of order $eT$), in addition to the nlo term. The leading order term must be subtracted as a counter-term to obtain the numerical result for the nlo contribution.

(3) For the real part of the electron self-energy, the (1-terms) have been calculated in \cite{Mitra} and shown to be of order $e^3T{\rm ln}(1/e),$ and not of order $e^2T$. In addition, we have shown by explicit calculation that the contribution from (2-terms) is also of order $e^3T{\rm ln}(1/e)$.  These results appear to indicate that the power counting argument over estimates the contribution from (1-terms) and (2-terms) terms. A similar result was obtained by Schulz in \cite{SCH}: he found that the contribution to the gluonic plasma frequency from 2-loop diagrams with hard internal momentum is smaller than predicted by power counting. It is not clear if (2-terms) terms are always over estimated by the power counting argument. This issue is discussed in Appendix E.

The ${\rm ln}(1/e)$ part of the contribution discussed above comes from an integral of the form
\bea
\int dp\;p\; \;(n_b(p)+n_f(p))\;{\rm Prin}\frac{1}{p^2-q_0^2}
\eea
where the notation ``${\rm Prin}$'' indicates a principle part and the factors $n_b(p)$ and $n_f(p)$ are bose-einstein and fermi-dirac distribution functions. The integral is cut off at large $T$ by the distribution functions and regulated at small $p$ by the soft  cutoff $q_0$. The point is that both of the distribution functions contribute to the result. This is not the case for the nlo contribution to the imaginary part which is dominated by the soft part of the momentum integral and the infra-red part of the bose-einstein distribution function. The result of the calculation in \cite{KKM,BPquarkdamp} justifies dropping all terms proportional to the fermi-dirac distribution function in the integrand for the imaginary part. In the expression for the real part we keep all terms containing either bose-eintein or fermi-dirac distributions.

(4) The numerical calculation of the real part of the self-energy is also more complicated than the corresponding integral for the imaginary part. The integrand for the real part contains a cut that gives large positive and negative contributions as it is approached from either side. These limits must be handled carefully to extract the surviving finite contribution. 

A compact analytic result for the real part of the fermion self-energy at nlo is given in Eqn. (\ref{real-part}). The result of the numerical calculation of this integral will be presented in a future publication.\\

This paper is organized as follows. In section \ref{prelim} we define some notation and discuss the diagrams that contribute to the fermion self-energy at nlo. In section \ref{CTP} we give some information about the technical aspects of the finite temperature calculation.  In section \ref{PROPS} we define our notation for the propagators and vertices. In section \ref{CALC} we describe the structure of the calculation and give the analytic results for the integrands of the retarded self-energy, and its real and imaginary parts. In section \ref{NUM} we give the numerical result for the imaginary part. In section \ref{CONC} we present our conclusions. In Appendix A we explain the notation we use to assign indices to the components of the Keldysh $n$-point functions and vertices. In Appendix B we give the full expressions for the discontinuities of the propagators. In Appendix C we list the ward identities and kms conditions that are needed to do the calculation. In Appendix D we give some explicit results for the vertex functions that cannot be rewritten using ward ientities. In Appendix E we discuss power counting arguments.

\section{Preliminaries}
\label{prelim}

We use capital letters to denote 4-momenta: $K = (k_0,\vec{k})$. We take the external momentum to be $Q = (q_0,\vec{0})$ and write $R=P+Q$ so that we have $\vec{r}=\vec{p}$. The fermion and photon thermal masses are: $m_f^2 = e^2 T^2/8$ and $m_G^2 = e^2T^2/6$. We use:
\bea
\{\gamma_\mu,\gamma_\nu\} = 2g_{\mu\nu}\;;~~g_{\mu\nu} = {\rm diag}\;(1,-1,-1,-1)\,.\nonumber
\eea
The thermal distribution functions are defined as:
\bea
\label{therm}
n_b(p) = \frac{1}{e^{\beta p}-1}\,;~~n_f(p) = \frac{1}{e^{\beta p}+1}\,;~~N_B(p) = 1+2n_b(p)\,;~~N_F(p) = 1-2n_f(p)\,.
\eea
In this paper we are only interested in thermal effects and consequently we ignore zero temperature pieces of the self-energy. 
The fermion self-energy is decomposed in the usual way:
\bea
{\bf \Sigma} = \gamma^0\Sigma^{(0)} + \vec{\gamma}\cdot\hat{q}\Sigma^{(i)}\,.
\eea
Since we have taken $q=0$ the only non-zero component is 
$\Sigma^{(0)} = {\rm Tr}\,(\gamma^0 {\bf \Sigma})/4$. 
From now on, we suppress the superscript `$(0)$' to simplify the notation. 
We use throughout the conventions of \cite{LeBellac}, but we work in real time.

The diagrams to be calculated are shown in Fig. 1, where dotted lines indicate photons and solid lines are fermions. The notation is defined in Fig. 2 where $P_{\psi {\rm in}}$ indicates the momentum of an incoming fermion, $P_{\psi {\rm out}}$ is the momentum of an outgoing fermion, and $P^\mu_{\gamma{\rm in}}$ is the momentum of an incoming photon. The vertices are given by: 
\bea
\label{vertex}
&& \Gamma_\mu(P_{\psi {\rm in}},P_{\psi {\rm out}}) = \gamma_\mu+\Gamma{\rm htl}_\mu(P_{\psi {\rm in}},P_{\psi {\rm out}})\\
&&M_{\mu\nu}(P_{\psi {\rm in}},P^\mu_{\gamma{\rm in}},P^\nu_{\gamma{\rm in}},P_{\psi {\rm out}})=M{\rm htl}_{\mu\nu}(P_{\psi {\rm in}},P^\mu_{\gamma{\rm in}},P^\nu_{\gamma{\rm in}},P_{\psi {\rm out}}) \nonumber
\eea
The htl vertex functions can be written as the contraction of a gamma matrix and a factor that does not contain dirac structure. We introduce the notation:
\bea
\label{htl-vertex}
&&\Sigma{\rm htl}:=\gamma^\lambda\Sigma{\rm htl}_\lambda \\
&&\Gamma{\rm htl}_\mu:=\gamma^\lambda \Gamma{\rm htl}_{\mu\lambda}\nonumber\\
&&M{\rm htl}_{\mu\nu}:=\gamma^\lambda M{\rm htl}_{\mu\nu\lambda} \nonumber
\eea
In order to keep the notation simple, we will not add a tilde to the vertices on the rhs of these expressions. The extra lorentz index indicates that the gamma matrix has been factor out. 
\par\begin{figure}[H]
\begin{center}
\includegraphics[width=10cm]{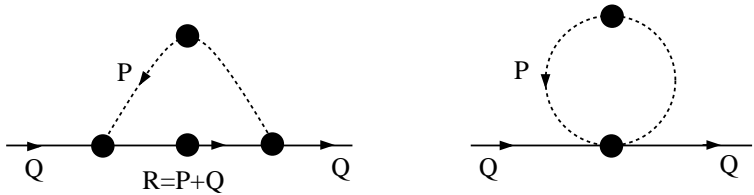}
\end{center}
\label{KKMdiag2}
\caption{The diagrams to calculate for a soft electron.} 
\end{figure}
\par\begin{figure}[H]
\begin{center}
\includegraphics[width=9cm]{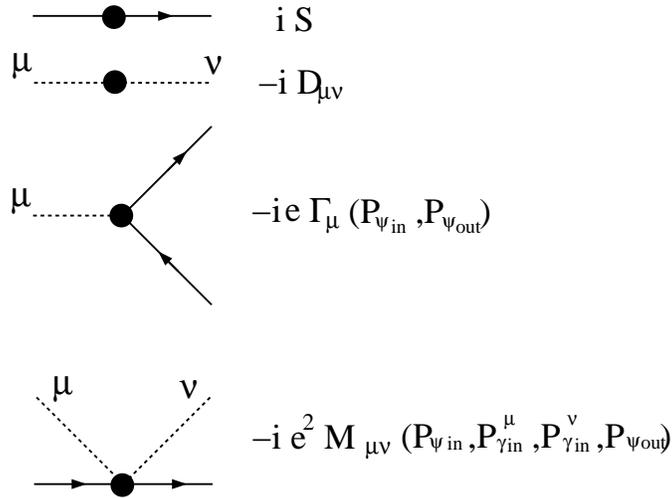}
\end{center}
\label{notation-diag}
\caption{Definitions of conventions for propagators and vertices.}
\end{figure}

All of the contributions identified by the power counting arguments of Braaten and Pisarski as contributing at nlo, as well as the leading-order contributions, are formally included in Fig. 1 if the integral over the internal momentum variable is not restricted to any particular region of the phase space, but runs over the full range of the momentum variable.

\section{Real Time Finite Temperature Field Theory in the Keldysh Representation}
\label{CTP}

We will do the calculation in the Keldysh representation of the real time formulation of finite temperature field theory. The basic structure of the method is described below for a scalar theory. Details are given in \cite{mcTF}. In the closed time path (CTP) formulation of finite temperature field theory, the path integral is defined on a contour in complex time that has two branches. The `1' branch runs from minus infinity to positive infinity just above the real axis, and the `2' branch runs back from positive infinity to negative infinity just below the real axis. 
\par\begin{figure}[H]
\begin{center}
\includegraphics[width=5cm]{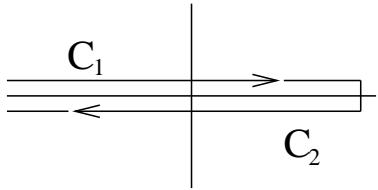}
\end{center}
\label{fig-contour}
\caption{The CTP contour in the complex time}
\end{figure}
\xx Green functions are defined on this contour:
\bea
iD_C (x,y)  
= \langle  \tilde T_c \phi (x) \phi (y) \rangle \nonumber
\eea
where $\tilde T_c$ indicates time ordering along the contour.
In the 1-2 representation, a propagator with real time arguments is written as  a 2$\times$2 matrix:
\bea
&&D = \left(\begin{array}{cc} D_{11} & D_{12} \\ D_{21} & D_{22} \end{array} \right)\\[2mm]
&&iD_{11}(x,y)  =   
\langle  T \phi (x) \phi (y) \rangle  \nonumber\\
&&iD_{12}(x,y)  = 
\langle  \phi (y) \phi (x) \rangle  \nonumber\\
&&iD_{21}(x,y)   = 
 \langle  \phi (x) \phi (y) \rangle \nonumber \\
&& iD_{22}(x,y)  = 
\langle  T^* \phi (x) \phi (y) \rangle \nonumber
\eea
where  the operators $T$ and $T^*$ indicate time ordering and anti-time ordering, respectively. Note that there are at most three independent components because of the constraint
\bea
\label{circ}
D_{11}-D_{12}-D_{21}+D_{22} = 0.
\eea

In general, in the 1-2 representation, an $n$-point function is a tensor with $2^n$ components, corresponding to the two possible values of each index. Only $2^n-1$ of these components are independent because of a constraint of the same form as the one given above. Truncating external legs gives the corresponding vertex functions. 
There are $(2^2-1=3)$ self-energy functions, $(2^3-1=7)$ 3-point vertices and $(2^4-1=15)$ 4-point vertices. 

In the Keldysh representation, we use linear combinations of components in the 1-2 representation that have a more direct physical interpretation. For example, the three components of the propagator in the Keldysh representation correspond to the retarded, advanced, and symmetric propagators.
For any Keldysh vertex function, each of the $2^n-1$ components is denoted by a single index which takes values $\{2,3,4,\cdots 2^n\}$. The definitions we use for assigning indices to the Keldysh functions are given in Appendix A.
For clarity, where possible, we will write propagators and self energies as retarded, advanced, or symmetric functions instead of using the corresponding numerical index.

We give explicitly below the expressions for the three propagator components, and the retarded and advanced self energies. The retarded, advanced and symmetric propagators are given by:
\bea
&& D_{ret}(K) = D_{11}-D_{12} = \frac{1}{K^2+i {\rm Sign}(k_0) \epsilon}\\
&& D_{adv}(K) = D_{11}-D_{21} = \frac{1}{K^2-i {\rm Sign}(k_0) \epsilon}\nonumber\\[2mm]
&& D_{sym}(K) = D_{11}+D_{22} = N_B(k_0)(D_{ret}(K)-D_{adv}(K))= -2\pi iN_B(k_0){\rm Sign}(k_0)\,\delta(K^2)\,. \nonumber
\eea
The retarded and advanced self energies (which give the real and imaginary parts: ${\rm Re}\,\Sigma_{ret} = 1/2(\Sigma_{ret}+\Sigma_{adv})\;;~~{\rm Im}\,\Sigma_{ret} = 1/(2i)(\Sigma_{ret}-\Sigma_{adv})$) are:
\bea
\Sigma_{ret} = \Sigma_{11}+\Sigma_{12}\,;~~\Sigma_{adv} = \Sigma_{11}+\Sigma_{21}\,.
\eea
The calculation of the diagrams in Fig. 1 is performed as follows. Each line carries an index on each end which can take values 1 or 2. One sums over all internal indices and takes the combinations of external indices that corresponds to the component of the self-energy that is wanted. The result is rotated to the Keldysh basis. The technique is described in \cite{mcTF}.

\section{htl propagators and vertices}
\label{PROPS}

The basic structure of QED at finite temperature is the same as for scalar theory, but we have to include the dirac and lorentz structure of the propagators and vertices. 

\subsection{Propagators}
In Eqns. (\ref{ferm1}), (\ref{ferm2}) and (\ref{photon-prop}), we simplify the notation by supressing the subscripts $ret$, $adv$. 
For photons the symmetric propagator is given by $D_{sym}^{\mu\nu}(P) = N_B(p_0)(D_{ret}^{\mu\nu}(P)-D_{adv}^{\mu\nu}(P))$ and for fermions we have $S_{sym}(R) = N_F(r_0)(S_{ret}(R)-S_{adv}(R))$.

The fermion propagator is written
\bea
\label{ferm1}
S(R) = i(\Rsl-\Sigma{\rm htl}(R))^{-1}
\eea
The inverse propagators are inverted by writing
\bea
\label{ferm2}
&&S(R)=\frac{1}{2}(S_+(R)(\gamma_0-\rhatsl)+\frac{1}{2}S_-(R)(\gamma_0+\rhatsl))  \\
&&S_+(R) = -\frac{2 r^2}{2 r \left(m_f^2+r
   \left(r-r_0\right)\right)+\ln \left(\frac{r_0+r}{r_0-r}\right) \left(r-r_0\right) m_f^2} \nonumber\\
&& S_-(R) = \frac{2 r^2}{2 r \left(m_f^2+r \left(r+r_0\right)\right)-\ln \left(\frac{r_0+r}{r_0-r}\right) m_f^2
   \left(r+r_0\right)}\nonumber
\eea
We use the covariant gauge and write the photon propagator in terms of transverse and longtitudinal projections (recall that $p=r$):
\bea
\label{photon-prop}
&&D_{\mu\nu}(P)=P_{\mu\nu}^T D_T(P) +P_{\mu\nu}^L \frac{p^2}{P^2}D_L(P)-\alpha\frac{P_\mu P_\nu}{P^2} \\[2mm]
&&D_T(P)=\frac{1}{P^2-G(p_0,r)}\,;~~D_L(P)=\frac{P^2}{r^2}\frac{1}{P^2-F(p_0,r)}\nonumber\\[2mm]
&& G(p_0,r) = \frac{1}{r^2}\left(1-\frac{{\cal Q}_0\left(p_0,r\right) p_0}{r}\right) P^2 m_G^2+m_G^2;~~~ {\cal Q}_0(p_0,r) = \frac{1}{2} \ln \left(\frac{p_0+r}{p_0-r}\right)\nonumber\\[2mm]
&& F(p_0,r) = -\frac{1}{r^2}2 m_G^2 \left(1-\frac{{\cal Q}_0\left(p_0,r\right) p_0}{r}\right) P^2\nonumber\\[2mm]
&& P_{\mu\nu}^T = g_{\mu  \nu }-\frac{Q^{\mu } Q^{\nu }}{q_0^2}+\frac{1}{r^2}\left(P^{\mu }-\frac{p_0 Q^{\mu }}{q_0}\right) \left(P^{\nu
   }-\frac{p_0 Q^{\nu }}{q_0}\right)\nonumber\\[2mm]
&& P_{\mu\nu}^L =-\frac{P^{\mu } P^{\nu }}{P^2}+\frac{Q^{\mu } Q^{\nu }}{q_0^2}-\frac{1}{r^2}\left(P^{\mu }-\frac{p_0 Q^{\mu}}{q_0}\right) \left(P^{\nu }-\frac{p_0 Q^{\nu }}{q_0}\right) \nonumber
\eea   
The expressions for the projection operators have been obtained by writing the vector that defines the rest frame of the medium as $n_\mu = \{1,0,0,0\} = Q_\mu/q_0$. 

The discontinuity of a scalar propagator is defined as:
\bea
\rho(K) = -i\, d(K) = -i(D_{ret}(K)-D_{adv}(K))
\eea
The discontinuities of the fermion and photon propagators are obtained from:
\bea
&&d_+(R) = S^+_{ret}(R)-S^+_{adv}(R)\,;~~d_-(R) = S^-_{ret}(R)-S^-_{adv}(R)\\[2mm]
&&d_T(P) = D^T_{ret}(P)-D^T_{adv}(P)
\,;~~d_L(P) = D^L_{ret}(P)-D^L_{adv}(P))\nonumber\,.
\eea
Each discontinuity has a pole and a cut contribution and has the general form:
\bea
d_x(K)=-2\pi i \sum_{n=\pm 1} Z_x(k)\; n\; \delta(k_0-n \,\omega_x(k))-2\pi i \beta_x(k_0,k)\,;~~~x=\{T,L,+,-\}
\eea
The functions $\omega_T(p)$, $\omega_L(p)$, $\omega_+(p)$ and $\omega_-(p)$ are the roots of the denominators of the corresponding propagators. The expressions for these functions, and for the residues $Z_i$ and cut functions $\beta_i$ are given in Appendix B.

\section{Calculation of the htl corrected 1-loop diagrams}
\label{CALC}

\subsection{structure of the calculation}

The first step in the calculation of the diagrams in Fig. 1 is to sum over indices in the 1-2 representation and rotate the result to the Keldysh basis. We define the operator 
\bea
\label{Nhat}
\hat N:= -\frac{i e^2}{32 \pi ^3} \int dp_0\int dr\,.
\eea
which will be factored out of all expressions. 
We use the techniques of \cite{mcTF} to obtain:
\bea 
&& \label{diag1} \Sigma_{ret}(Q)\Big|_{{\rm diagram~1}} \\
&&= \hat N \;r^2\;{\rm Tr}\Big[\Big.\gamma^0 D^{\text{sym}}_{\mu\nu}(P) \Gamma ^{\nu
   }[\,2\,](Q,R) S_{ret}(R)  \Gamma ^{\mu }[\,2\,](R,Q) 
   + D^{adv}_{\mu\nu}(P)\Gamma ^{\nu
   }[\,2\,](Q,R)S_{\text{sym}}(R) \Gamma ^{\mu }[\,3\,](R,Q)  \nonumber\\
&& %
  ~~~~~~ +D^{adv}_{\mu\nu}(P)  \Gamma ^{\nu
   }[\,2\,](Q,R)S_{ret}(R) \Gamma ^{\mu }[\,4\,](R,Q)
+D^{ret}_{\mu\nu}(P)  \Gamma ^{\nu
   }[\,4\,](Q,R) S_{ret}(R) \Gamma ^{\mu }[\,2\,](R,Q) \nonumber \\
&&~~~~~~+ D^{adv}_{\mu\nu}(P)  \Gamma ^{\nu
   }[\,6\,](Q,R)S_{adv}(R)\Gamma ^{\mu }[\,3\,](R,Q)\Big]\nonumber\\[4mm]
&&\label{diag2} \Sigma_{ret}(Q)\Big|_{{\rm diagram~2}}  \\
&& = \hat N \;\frac{1}{2}\,r^2\;{\rm Tr}\Big[\Big.\gamma^0 
\left(D^{\text{sym}}_{\mu\nu}(P) M^{\mu\nu}[\,2\,](Q,P,-P,Q)+D^{ret}_{\mu\nu}(P) M^{\mu\nu}[\,4\,](Q,P,-P,Q)+D^{adv}_{\mu\nu}(P) M^{\mu\nu}[\,6\,](Q,P,-P,Q)\right)\Big]\nonumber
\eea
The indices in square brackets refer to the Keldysh components of the vertices. The notation is defined in Appendix A.

The basic strategy of our calculation is to rewrite the htl vertices in terms of the self energies, and rearrange the result to obtain the simplest possible form. In order to do this, we first note that there are four basic structures in the transverse and longtitudinal projection operators (\ref{photon-prop}):
\bea
\label{structure}
P_\mu P_\nu\,;~~\Big\{\frac{Q_\mu}{q_0} P_\nu,\frac{Q_\nu}{q_0} P_\mu\Big\}\,;~~\frac{Q_\mu Q_\nu}{q_0^2}\,;~~g_{\mu\nu}\,. 
\eea

The first two forms give a zero contribution to the final result. This is a consequence of gauge invariance and can be derived using the ward identities. One finds that the gauge dependent contribution to the nlo fermion self-energy is proportional to an integral times the square of the inverse propagator $S^{(-1)}(Q) = \Qsl - \Sigma{\rm htl}(Q),$ which vanishes on the mass shell. In \cite{BKS} it was pointed out that a straightforward evaluation of the integral produces mass-shell singularities that cancel the contributions from the two inverse propagators and give a finite gauge dependent contribution to the damping rate. This problem was resolved by Rebhan \cite{Tonycomment} who showed that the integral must be regulated before the mass shell is approached. Using this procedure one finds that the position of the pole is gauge independent, and the gauge dependence occurs only in the unphysical residue.  

The third and fourth forms in (\ref{structure}) give non-zero contributions. All terms that come from the third form ($Q_\mu Q_\nu/q_0^2$) can be rewritten in terms of the fermion self-energy by using the htl ward identities. Terms proportional to $g_{\mu\nu}$ are more difficult to handle. We use the kms conditions for the 3-point functions, and some equations obtained by taking the complex conjugate and interchanging the arguments of the fermion legs, to obtain the simplest possible expression. A complete list of the ward identities and kms conditions for 3- and 4-point functions are found in \cite{mcTF}. In Appendix C we have collected the ones that we use in this calculation. 

In order to express the results in a compact form, we rewrite the photon propagator in (\ref{photon-prop}) keeping only the terms that will give non-zero contributions:
\bea
\label{new-photon-prop}
&& D'_{\mu\nu}(P) = g_{\mu\nu}D_T(P) + \frac{Q_\mu Q_\nu}{q_0^2}(\tilde D_T(P)+\tilde D_L(P))\\
&& \tilde D_T(P)= \frac{P^2}{p^2}D_T(P)\,;~~\tilde D_L(P)= -D_L(P)\,.\nonumber
\eea
We will use the same notation for the discontinuities as before: $\tilde d_{T/L}(P) = \tilde D^{ret}_{T/L}(P)-\tilde D^{adv}_{T/L}(P)$. Using the fact that the htl 4-point vertex satisfies $M{\rm htl}_\mu^{~\mu}=0$, we consider the structure of the terms that result when we use (\ref{new-photon-prop}) in (\ref{diag1}) and (\ref{diag2}).   There are three types of terms:\\

\xx (1) contributions from the first term in (\ref{new-photon-prop})\\

\xx (2) contributions from the second and third terms in (\ref{new-photon-prop}) which can be divided into two types:

(2a) terms which do not contain fermion propagators 

(2b) terms that do contain fermion propagators\\

\xx All contributions to $\Sigma_{ret}(Q)$ have one of the forms below:
\bea
\label{type-1}
&&(1) = \hat N\;r^2 d_{T}(P) N_B\left(p_0\right){\rm Tr}(\gamma^0\gamma^{\tau} S^\pm_{ret}(R)\gamma^{\lambda}) \Gamma_{\tau\mu}[\,2\,](Q,R)\, \Gamma_{\lambda}^{~\mu}[\,2\,](Q,R)\\
&&~~~~+ \hat N\; r^2 D_T^{adv}(P) N_F\left(r_0\right) {\rm Tr}\Big[\gamma^0\gamma^{\tau}\Big(S^\pm_{ret}(R) \Gamma_{\tau\mu}[\,2\,](Q,R)\, \Gamma_{\lambda}^{~\mu}[\,2\,](Q,R)  -S^\pm_{adv}(R) \Gamma_{\tau\mu}[\,5\,](Q,R)\, \Gamma_{\lambda}^{~\mu}[\,5\,](Q,R)\Big)\gamma^{\lambda} \Big]  \nonumber\\[2mm]
&&(2a)=-\frac{1}{q_0^2} \hat N\;r^2 \tilde d_{T/L}(p) {\rm Tr}(\gamma^0\gamma^\tau) \left(2 Q_\tau+R_\tau\right) N_B\left(p_0\right)\label{2a}\\
&&(2b) = \frac{1}{q_0^2} \hat N\;r^2 \left(Q_{\tau}+R_{\tau}\right) {\rm Tr}\left[\gamma^0 \gamma^{\tau}\left(\tilde d_{T/L}(P) S^\pm_{ret}(R)
   N_B\left(p_0\right)+\tilde D^{adv}_{T/L}(P) d^\pm(R) N_F\left(r_0\right)\right)\gamma^{\lambda}\right]\left(Q_{\lambda}+R_{\lambda}\right) \label{2b}
\eea
Using Eqns. (\ref{vertex}) and (\ref{htl-vertex}) we have written: 
\bea
\Gamma_{\tau\mu}[\,i\,](Q,R) = g_{\tau\mu}+\Gamma{\rm htl}_{\tau\mu}[\,i\,](Q,R)\,;~~i=2~{\rm or}~5\,.
\eea
There are two contributions from terms of type-(1) which come from the (+/ -) modes of the electron. There are two contributions from terms of type-(2a) which come from transverse and longtitudinal photon modes. 
There are four type-(2b) contributions which come from the (transverse/longtitudinal) $\times$ (+/ -) modes of the photon and electron.

\subsection{Longitudinal results}

Longtitudinal modes get contributions from terms of type-(2a) and type-(2b) only. For future convenience, we also divide results into contributions that depend on boson distribution functions and contributions that depend on fermion distribution functions. The contributions to $\Sigma_{ret}(Q)$ from longtitudinal photon modes are:
\bea
\label{long-res}
&&\text{Type2a}\left(\text{Long},N_B\right)=\frac{12}{q_0} \,\hat N\, r^2\,d_L(P) N_B\left(p_0\right)\\
&& \text{Type2a}\left(\text{Long},N_F\right)=0 \nonumber\\
&&\text{Type2b}\left(\text{Long},+,N_B\right)=-\frac{2}{q_0^2} \,\hat N\, r^2\, \left(-r+q_0+r_0\right)^2 d_L(P)
   S^{+}_{ret}(R) N_B\left(p_0\right)\nonumber\\
&&\text{Type2b}\left(\text{Long},-,N_B\right)=-\frac{2}{q_0^2} \,\hat N\, r^2\,\left(r+q_0+r_0\right)^2 d_L(P)
    S^{-}_{ret}(R)N_B\left(p_0\right)\nonumber\\
&&\text{Type2b}\left(\text{Long},+,N_F\right)=-\frac{2}{q_0^2}\,\hat N\, r^2\, \left(-r+q_0+r_0\right)^2 a_L(P)
   d_{+}(R) N_F\left(r_0\right)\nonumber\\
&&\text{Type2b}\left(\text{Long},-,N_F\right)=-\frac{2}{q_0^2}\,\hat N\, r^2\, \left(r+q_0+r_0\right)^2 a_L(P)
   d_{-}(R) N_F\left(r_0\right)\nonumber
\eea

\subsection{Transverse results}

Transverse modes also contain contributions from terms of type-(1). These terms can be further simplified by using explicit results for the htl 3-point vertex functions, which have a particularly simple form when one of the fermions is not moving. The expressions that we need are given in Appendix D.  
Using these results, all components of the vertices in (\ref{type-1}) can be written as simple functions of the htl self-energy. 
These self-energies also appear in the denominators of the htl fermion propagators. The general strategy is to rearrange terms in the numerators so that we can cancel as many terms as possible with the corresponding terms in the denominators. Significant simplifications occur when we combine terms and use the mass shell condition $q_0^2=m_f^2$. We define the notation:
\bea
\text{$\Sigma $htl}_0(s)(R) =\text{$\Sigma $htl}^{ret}_0(R) +\text{$\Sigma $htl}^{adv}_0(R)\,;~~\text{$\Sigma $htl}_0(d)(R) =\text{$\Sigma $htl}^{ret}_0(R) -\text{$\Sigma $htl}^{adv}_0(R)\,.
\eea
The contributions to $\Sigma_{ret}(Q)$ from transverse photon modes are:
\bea
\label{trans-res}
&&\text{Type2a}\left(\text{Trans},N_B\right)=-\frac{2}{q_0^2}\,\hat N\, d_T(P) N_B\left(p_0\right) \left(-3 r_0 R^2+4 m_f^2
   r_0+8 r^2 q_0+6 P^2 q_0-R^2 \text{$\Sigma $htl}^{ret}_0(R)\right) \nonumber\\
&& \text{Type2a}\left(\text{Trans},N_F\right)=\frac{2}{q_0^2}\,\hat N\, \left(r_0^2-r^2\right) a_T(P) N_F\left(r_0\right)
   \text{$\Sigma $htl}_0(d)(R) \\
&& \text{Type2b}\left(\text{Trans},+,N_B\right)=-\frac{1}{q_0^2}\,\hat N\,\left(p_0-r\right)^2 \left(r+p_0+2 q_0\right)^2
   d_T(P) N_B\left(p_0\right) S^{+}_{ret}(R) \nonumber \\
&& \text{Type2b}\left(\text{Trans},-,N_B\right)=-\frac{1}{q_0^2}\,\hat N\,\left(r+p_0\right)^2 \left(-r+p_0+2 q_0\right)^2
   d_T(P) N_B\left(p_0\right) S^{-}_{ret}(R)\nonumber\\
&& \text{Type2b}\left(\text{Trans},+,N_F\right)=-\frac{1}{q_0^2}\,\hat N\,\left(p_0-r\right)^2 \left(r+p_0+2 q_0\right)^2
   a_T(P) d_{+}(R) N_F\left(r_0\right) \nonumber\\
&& \text{Type2b}\left(\text{Trans},-,N_F\right)=-\frac{1}{q_0^2}\,\hat N\,\left(r+p_0\right)^2 \left(-r+p_0+2 q_0\right)^2
   a_T(P) d_{-}(R) N_F\left(r_0\right) \nonumber
\eea

\subsection{Imaginary parts}
The damping rate is determined from the imaginary parts of the expressions in (\ref{long-res}) and (\ref{trans-res}). The nlo contribution comes only from terms that depend on the bose-einstein distribution function. The contributions to ${\rm Im}\Sigma_{ret}(Q)$ are:
\bea
\label{imag-part}
&& \text{Im}(L,+,2 b)=\frac{i}{ q_0^2}\;\hat N \; r^2 \left(-r+q_0+r_0\right)^2 d_L(P) d_{+}(R)
   N_B\left(p_0\right) \\
&& \text{Im}(L,-,2 b)=\frac{i}{q_0^2}\;\hat N \; r^2 \left(r+q_0+r_0\right)^2 d_L(P) d_{-}(R)
   N_B\left(p_0\right) \nonumber\\
&& \text{Im}(T,2 a)=-\frac{i}{q_0^2}\;\hat N \; R^2 d_T(P) N_B\left(p_0\right) \text{$\Sigma $htl}_0(d)(R) \nonumber\\
&& \text{Im}(T,+,2 b)=\frac{i}{2q_0^2}\;\hat N \;\left(p_0-r\right)^2 \left(r+p_0+2 q_0\right)^2 d_{+}(R) d_T(P)
   N_B\left(p_0\right)\nonumber\\
&& \text{Im}(T,-,2 b)=\frac{i}{2q_0^2}\;\hat N \;\left(r+p_0\right)^2 \left(-r+p_0+2 q_0\right)^2 d_{-}(R) d_T(P)
   N_B\left(p_0\right) \nonumber
\eea
\subsection{Real parts}
As discussed in the introduction, the real parts could receive contributions from terms with both bose-einstein and fermi-dirac distribution functions. In addition, there are tadpole-type terms that contribute to the real part. We use the notation ``${\rm Prin}$'' to indicate a principle part. The contributions to ${\rm Re}\Sigma_{ret}(Q)$  are:
\bea
\label{real-part}
&& \text{Re}(L,N_B,2 a)=\frac{12}{q_0} \;\hat N \;r^2 d_L(p) N_B\left(p_0\right)\\
&& \text{Re}(L,+,N_B,2 b)=-\frac{2}{q_0^2} \;\hat N \;r^2 \left(-r+q_0+r_0\right)^2 d_L(P) N_B\left(p_0\right)
   \text{Prin}_{+}(R)\nonumber\\
&& \text{Re}(L,-,N_B,2 b)=-\frac{2}{q_0^2} \;\hat N \;r^2 \left(r+q_0+r_0\right)^2 d_L(P) N_B\left(p_0\right)
   \text{Prin}_{-}(R) \nonumber\\
&& \text{Re}(L,+,N_F,2 b)=-\frac{2}{q_0^2}\;\hat N \; r^2 \left(-r+q_0+r_0\right)^2 d_{+}(R)
   N_F\left(r_0\right) \text{Prin}_L(P) \nonumber\\
&& \text{Re}(L,-,N_F,2 b)=-\frac{2}{q_0^2} \;\hat N \;r^2 \left(r+q_0+r_0\right)^2 d_{-}(R)
   N_F\left(r_0\right) \text{Prin}_L(P) \nonumber\\
&& \text{Re}(T,N_B,2 a)=-\frac{1}{q_0^2}\;\hat N \; d_T(p) N_B\left(p_0\right) \left(2 \left(-3 r_0 R^2 +4 m_f^2 r_0+8 r^2 q_0+6 P^2 q_0\right)-R^2 \text{$\Sigma $htl}_0(s)(R)\right)\nonumber\\
&& \text{Re}(T,N_F,2 a)=\frac{2}{q_0^2}\;\hat N \; R^2 N_F\left(r_0\right) \text{Prin}_T(P) \text{$\Sigma
   $htl}_0(d)(R) \nonumber\\
&& \text{Re}(T,+,N_F,2 b)=-\frac{1}{q_0^2} \;\hat N \;\left(p_0-r\right)^2 \left(r+p_0+2 q_0\right)^2 d_T(P)
   N_B\left(p_0\right) \text{Prin}_{+}(R)\nonumber\\
&& \text{Re}(T,-,N_B,2 b)=-\frac{1}{q_0^2} \;\hat N \;\left(r+p_0\right)^2 \left(-r+p_0+2 q_0\right)^2 d_T(P)
   N_B\left(p_0\right) \text{Prin}_{-}(R)\nonumber\\
&& \text{Re}(T,+,N_F,2 b)=-\frac{1}{q_0^2} \;\hat N \;\left(p_0-r\right)^2 \left(r+p_0+2 q_0\right)^2 d_{+}(R)
   N_F\left(r_0\right) \text{Prin}_T(P)\nonumber\\
&& \text{Re}(T,-,N_F,2 b)=-\frac{1}{q_0^2}\;\hat N \; \left(r+p_0\right)^2 \left(-r+p_0+2 q_0\right)^2
   d_{-}(R) N_F\left(r_0\right) \text{Prin}_T(P) \nonumber
\eea

\section{Numerical evaluation of the Imaginary Part}
\label{NUM}

The damping rate can be calculated from the imaginary part of the self-energy using (\ref{damp}). 
Each of the four type-(2b) terms in (\ref{imag-part}) contains two discontinuities in the form of a product $d_{T/L}(P) d_{\pm}(R)$. Each of these discontinuities contains an cut piece and a pole piece, as given in Eqns. (\ref{cutpole-defn}), (\ref{cut}), (\ref{residue}) and (\ref{pole}) in Appendix B. The type-(2a) term in (\ref{imag-part}) contains one discontinuity of the form $d_T(P)$ which has a cut piece and a pole piece. Thus, the total number of terms to be calculated is: $(4\times 4)+(1\times 2) = 18$. 

We scale all dimensionful parameters with the leading order fermion mass: $r_0\rightarrow m_f r_0$; $r \rightarrow  m_f r$; $m_G^2 = x m_f^2$. The result for the quark damping rate for 2 flavours and 3 colours is:
\bea
&& x=\frac{(N_c+N_f/2)/6}{C_F/8}\,;~~C_F = (N_c^2-1)/(2N_c)\,;~~\{N_c=3,~~N_f=2\} ~\Rightarrow ~x=4 \\
&& \gamma_{QCD}(x=4) = \frac{g^2 T C_F}{4\pi}(2.81)\nonumber
\eea
This result agrees with that of \cite{KKM,BPquarkdamp}.

For the electron damping rate in a QED plasma we find:
\bea
x=m_G^2/m_f^2 = 4/3\,; ~~~ \Rightarrow~~~~ \gamma_{QED} = \frac{e^2 T}{4\pi}(2.70)
\eea

\section{Conclusions}
\label{CONC}

We have studied next-to-leading order contributions to the soft static fermion dispersion relation in hot QED. 
We  have derived an expression for the complete next-to-leading order contribution to the retarded fermion self-energy. The real and imaginary parts of this expression give the next-to-leading order contributions to the mass and damping rate of the fermionic quasi-particle. Many of the terms that are expected to contribute, according to the traditional power counting arguments, are actually subleading. We have explained why the power counting method over estimates the contribution from sub-leading pieces of the 1-loop diagram, and each of the 2-loop diagrams with both loop momenta hard. 

For the electron damping rate in QED we obtain: $\gamma_{QED} = \frac{e^2 T}{4\pi}(2.70)$. We have checked our method by calculating the next-to-leading order contribution to the  damping rate for the case of QCD with two flavours and three coulours. Our result agrees with the result obtained previously in the literature.
The numerical evaluation of the nlo contribution to the mass is left to a future publication.

\appendix
\section{Definitions of the Keldysh Vertices}\label{AppendixA}

As explained in section \ref{CTP}, green functions and vertices in the Keldysh representation are linear combinations of the green functions and vertices in the 1-2 representation. 
An $n$-point function in the 1-2
basis is written:
$G_{b_1,b_2,b_3,\cdots b_n}$ 
where the indices $b_i$ and take the values 1 or 2.
Vertices are obtained by truncating external legs:
\begin{equation}
\label{trun}
G_{b_1\cdots b_n}= G_{b_1 \bar b_1} \cdots G_{b_n \bar b_n} 
	\Gamma^{\bar b_1\cdots \bar b_n}  \,.
\end{equation}
Keldysh indices are written $\alpha_i$ and are assigned the values $\alpha = 1 := r$
and $\alpha = 2:= a$.
An $n$-point function in the Keldysh basis is written:
$G_{\alpha_1,\alpha_2,\alpha_3,\cdots \alpha_n}$. 
Vertices are obtained by truncating external legs:
\begin{equation}
\label{trun2}
G_{\alpha_1\cdots \alpha_n}= G_{\alpha_1 \bar \alpha_1} \cdots G_{\alpha_n \bar \alpha_n} 
	\Gamma^{\bar \alpha_1\cdots \bar \alpha_n}  \,.
\end{equation}

To rotate from the 1-2 representation to the Keldysh representation we use the transformation matrix:
\begin{equation}
\label{firstU}
U=\frac{1}{\sqrt{2}}\left(
\begin{array}{lr}

1 & 1 \\
1 & -1
\end{array}
\right).
\end{equation}
The $n$-point
function in the Keldysh representation is given by:
\begin{equation}
\label{Keldyn}
G_{\alpha_1\cdots\alpha_n}=2^{\frac{n}{2}-1}\, U_{\alpha_1}\!^{b_1} 
\cdots U_{\alpha_n}\!^{b_n} G_{b_1\cdots b_n}\,.
\end{equation}

In order to compactify the notation, we replace each string of indices $\{\alpha_1,\alpha_2,\cdots\}$ by a single numerical index:
\bea
G_{\alpha_1\alpha_2\cdots \alpha_n}: = G[\,i\,]\,;~~\Gamma^{\alpha_1\alpha_2\cdots \alpha_n}: = \Gamma[\,i\,]\,.
\eea 
We assign 
the choices of the variables $\alpha_1 \alpha_2 \cdots \alpha_n$ to the variable $i$ using the vector
\bea 
\label{ralist}
V_n = 
\Big(
\begin{array}{c}
r_n \\
a_n
\end{array}
\Big) \cdots \otimes
\Big(
\begin{array}{c}
r_2 \\
a_2
\end{array}
\Big)\otimes
\Big(
\begin{array}{c}
r_1 \\
a_1
\end{array}
\Big)
\eea
where the symbol $\otimes$ indicates the outer product. For each $n$,
the $i$th component of the vector corresponds to a list of variables
that is assigned the number $i$.  For clarity,
the results are listed below. To simplify the notation we drop the
subscripts and write a list like ``$r_1 r_2 a_3$'' as ``$rra$.'' \\

\xx [a] 2-point functions: $rr \ra 1$, $ar\ra 2$, $ra\ra 3$, $aa\ra 4$\\

\xx [b] 3-point functions: $rrr\ra 1$, $arr\ra 2$, $rar\ra 3$, $aar\ra 4$, $rra\ra 5$, $ara\ra 6$, $raa\ra 7$, $aaa\ra 
8$\\

\xx [c] 4-point functions: $rrrr \ra 1$, $arrr \ra 2$,  $ rarr \ra 3$,  $ aarr \ra 4$,  $ rrar \ra 5$,  $ arar \ra 6$,  $ raar \ra 7$,  $aaar \ra 8$,  $ rrra \ra 9$,  $ arra \ra 10$,  $ rara \ra 11$,  $ aara \ra 12$,  $ rraa \ra 13$,  $araa \ra 
14$,  $ raaa \ra 15$, $ aaaa \ra 16$\\

Using expressions like (\ref{circ}), it is easy to show that 
\bea
G_{aaaa\cdots a}=0 \,;~~\Gamma^{rrrr\cdots r}=0 ~~\rightarrow ~~ G[2^n]=0\,;~~\Gamma[1]=0\,.
\eea

\section{Discontinuities}\label{AppendixB}

Each discontinuity has a pole and a cut contribution and has the general form:
\bea
\label{cutpole-defn}
d_x(K)=-2\pi i \sum_{n=\pm 1} Z_x(k)\; n\; \delta(k_0-n \omega_x(k))-2\pi i \beta_x(k_0,k)\,;~~~x=\{T,L,+,-\}\,.
\eea
The functions $\omega_T(p)$, $\omega_L(p)$, $\omega_+(p)$ and $\omega_-(p)$ are the roots of the denominators of the corresponding propagators. The expressions for these functions, and for the residues $Z_i$ and cut functions $\beta_i$ are given below.

The cut functions are:
\bea
\label{cut}
&&\beta _T(p_0,r)=-\frac{2 r^3 m_G^2 p_0 \left(p_0^2-r^2\right) \Theta \left(r^2-p_0^2\right)}{\pi ^2 p_0^2 \left(r^2-p_0^2\right)^2 m_G^4+\left(2 r^5-2 p_0^2
   r^3+\ln \left|\frac{r_0+r}{r_0-r}\right| m_G^2 p_0 r^2+2 m_G^2 p_0^2 r-\ln \left|\frac{r_0+r}{r_0-r}\right| m_G^2
   p_0^3\right)^2} \\
&& \beta _L(p_0,r)=\frac{r m_G^2 p_0 \Theta \left(r^2-p_0^2\right)}{\pi ^2 p_0^2 m_G^4+\left(r^3+2 m_G^2 r-\ln \left|\frac{r_0+r}{r_0-r}\right| m_G^2
   p_0\right)^2} \nonumber\\
&& \beta _{+}(r_0,r)=\frac{2 r^2 m_f^2 \left(r-r_0\right) \Theta \left(r^2-r_0^2\right)}{\pi ^2 \left(r-r_0\right)^2 m_f^4+\left(2 \left(r-r_0\right)
   r^2+m_f^2 \left(r \left(\ln \left|\frac{r_0+r}{r_0-r}\right|+2\right)-\ln \left|\frac{r_0+r}{r_0-r}\right| r_0\right)\right)^2}\nonumber\\
&& \beta _{-}(r_0,r)=\frac{2 r^2 m_f^2 \left(r_0+r\right) \Theta \left(r^2-r_0^2\right)}{\pi ^2 \left(r+r_0\right)^2 m_f^4+\left(m_f^2 \left(r \left(\ln
   \left|\frac{r_0+r}{r_0-r}\right|-2\right)+\ln \left|\frac{r_0+r}{r_0-r}\right| r_0\right)-2 r^2 \left(r+r_0\right)\right)^2}\nonumber
\eea
The residues are:
\bea
\label{residue}
&&Z_T(r)=\frac{\omega _T(r) \left(r^2-\omega _T(r)^2\right)}{r^4+\omega _T(r)^4-2 \left(r^2+m_G^2\right) \omega
   _T(r)^2}\\
&& Z_L(r)=\frac{\omega _L(r)^3-r^2 \omega _L(r)}{r^2 \left(r^2+2 m_G^2-\omega _L(r)^2\right)} \nonumber\\
&& Z_{+}(r)=\frac{\omega _{+}(r)^2-r^2}{2 m_f^2} \nonumber\\
&& Z_{-}(r)=\frac{\omega _{-}(r)^2-r^2}{2 m_f^2} \nonumber
\eea
The positions of the poles are determined from the solutions of the equations:
\bea
\label{pole}
&& \ln \left|\frac{\omega _T(r)+r}{\omega _T(r)-r}\right| \omega _T(r) \left(r^2-\omega _T(r)^2\right)
   m_G^2+2 \left(r^5-\omega _T(r)^2 r^3+m_G^2 \omega _T(r)^2 r\right)=0\\
&&r^3+2 m_G^2 r-\ln \left|\frac{\omega _L(r)+r}{\omega _L(r)-r}\right| m_G^2 \omega _L(r)=0\nonumber \\
&& \ln \left|\frac{\omega _{+}(r)+r}{\omega _{+}(r)-r}\right| \left(r-\omega_{+}(r)\right) m_f^2+2 r \left(m_f^2+r \left(r-\omega _{+}(r)\right)\right)=0 \nonumber \\
&&2 r \left(m_f^2+r \left(r+\omega _{-}(r)\right)\right)-\ln \left|\frac{\omega_{-}(r)+r}{\omega _{-}(r)-r}\right| m_f^2 \left(r+\omega _{-}(r)\right)=0\nonumber
\eea

\section{Ward Identities and KMS Conditions}\label{AppendixC}

The ward identities at finite temperature have the same form as at zero temperature. As discussed in section \ref{CTP}, there are 7  components of the 3-point function and 15 components of the 4-point function, each of which has its own ward identity. These ward identities are expressed in a compact notation as:
\bea
&& (P_{\rm out}-P_{\rm in})^\mu \Gamma{\rm htl}_\mu[\,i\,](P_{\rm in},P_{\rm out}) =\Sigma{\rm htl}[\,j_1\,](P_{\rm in})-\Sigma{\rm htl}[\,j_2\,](P_{\rm out}) \\
&&  Q^\mu M{\rm htl}_{\mu\nu} [\,i\,](P_{\rm in},Q_1,Q_2,P_{\rm out}) = \Gamma{\rm htl}_\nu[\,j_1\,](P_{\rm in},P_{\rm out}-Q_1)-\Gamma{\rm htl}_\nu[\,j_2\,](P_{\rm in}+Q_1,P_{\rm out})\nonumber
\eea
The index in square brackets denotes the Keldysh component of the vertex. The complete sets of indices are given in \cite{mcTF}. \\

In the htl limit, the ward identities have a somewhat simpler structure. We give only the results that are needed for this calculation (using $S:=-P+Q$):
\bea
\begin{array}{ll}
Q_{\mu} \text{$\Gamma $htl}^{\mu \nu}[\,2\,](Q,R) =  \text{$\Sigma $htl}^{\nu }_{ret}(R)\,;~~ & R_\nu \text{$\Gamma $htl}^{\mu\nu}[\,2\,](Q,R)= \text{$\Sigma $htl}^{\mu }_{ret}(Q) \\
Q_\mu \text{$\Gamma $htl}^{\mu\nu}[\,4\,](Q,R) =  0 \,;~~& R_\nu \text{$\Gamma $htl}^{\mu\nu}[\,4\,](Q,R) = 0 \\
Q_\mu \text{$\Gamma $htl}^{\mu\nu}[\,6\,](Q,R) = \text{$\Sigma $htl}^{\nu}_{\text{sym}}(R) \,;~~ & R_{\nu } \text{$\Gamma $htl}^{\mu  \nu }[\,6\,](Q,R)=0 \\
Q_{\nu } \text{$\Gamma $htl}^{\mu  \nu }[\,2\,](R,Q)=\text{$\Sigma $htl}^{\mu }_{ret}(R)\,;~~ & R_{\mu } \text{$\Gamma $htl}^{\mu  \nu }[\,2\,](R,Q)=\text{$\Sigma $htl}^{\nu }_{ret}(Q)\\
Q_{\nu } \text{$\Gamma $htl}^{\mu  \nu }[\,3\,](R,Q)=\text{$\Sigma $htl}^{\mu }_{adv}(R)\,;~~ & R_{\mu } \text{$\Gamma $htl}^{\mu  \nu }[\,3\,](R,Q)=\text{$\Sigma $htl}^{\nu }_{ret}(Q)\\
Q_{\nu } \text{$\Gamma $htl}^{\mu  \nu }[\,4\,](R,Q)=\text{$\Sigma $htl}^{\mu }_{\text{sym}}(R)\,;~~ & R_{\mu } \text{$\Gamma $htl}^{\mu  \nu }[\,4\,](R,Q)=0
\end{array} 
\eea
\bea
&& Q_\mu Q_\nu M^{\mu\nu\tau}[\,2\,](Q,P,-P,Q) = \Sigma{\rm htl}^\tau_{ret}(R)+\Sigma{\rm htl}^\tau_{ret}(S)\\
&& Q_\mu Q_\nu M^{\mu\nu\tau}[\,4\,](Q,P,-P,Q) = N_F(s_0)(\Sigma{\rm htl}^\tau_{ret}(S)-\Sigma{\rm htl}^\tau_{adv}(S))\nonumber\\
&& Q_\mu Q_\nu M^{\mu\nu\tau}[\,6\,](Q,P,-P,Q) = N_F(r_0)(\Sigma{\rm htl}^\tau_{ret}(R)-\Sigma{\rm htl}^\tau_{adv}(R))\nonumber
\eea

Terms which are contracted with the part of the photon propagator that is proportional to $g_{\mu\nu}$ cannot be written in terms of the self-energy using ward identities. We simplify  these terms using the kms conditions for the 3-point functions: 
\bea
&&\text{$\Gamma $htl}_{\mu }[\,4\,](Q,R)\\
&&~~=-N_B\left(p_0\right) \text{$\Gamma $htl}_{\mu }[\,2\,](Q,R)-N_F\left(q_0\right)
   \text{$\Gamma $htl}_{\mu }[\,3\,](Q,R)+N_B\left(p_0\right) \text{$\Gamma $htl}^*_{\mu }[\,5\,](Q,R)+N_F\left(q_0\right)
   \text{$\Gamma $htl}^*_{\mu }[\,5\,](Q,R) \nonumber\\
&& \text{$\Gamma $htl}_{\mu }[\,6\,](Q,R)\nonumber\\
&&~~=N_F\left(r_0\right) \text{$\Gamma $htl}_{\mu }[\,2\,](Q,R)-N_F\left(q_0\right)
   \text{$\Gamma $htl}_{\mu }[\,5\,](Q,R)+N_F\left(q_0\right) \text{$\Gamma $htl}^*_{\mu
   }[\,3\,](Q,R)-N_F\left(r_0\right) \text{$\Gamma $htl}^*_{\mu }[\,3\,](Q,R)\nonumber\\
&&\text{$\Gamma $htl}_{\mu }[\,4\,](R,Q)\nonumber\\
&&~~= N_B\left(p_0\right) \text{$\Gamma $htl}_{\mu }(2)(R,Q)-N_F\left(r_0\right)
   \text{$\Gamma $htl}_{\mu }(3)(R,Q)-N_B\left(p_0\right) \text{$\Gamma $sthtl}_{\mu }(5)(R,Q)+N_F\left(r_0\right)
   \text{$\Gamma $sthtl}_{\mu }(5)(R,Q)\nonumber
\eea
Some additional relations between vertices are:
\bea
&&\text{$\Gamma $htl}^*_{\mu }[\,5\,](X,Y)= \text{$\Gamma $htl}_{\mu
   }[\,2\,](Y,X)\\
&&\text{$\Gamma $htl}^*_{\mu }[\,2\,](X,Y)=
 \text{$\Gamma $htl}_{\mu }[\,5\,](Y,X)\nonumber\\
&&\text{$\Gamma $htl}^*_{\mu}[\,3\,](X,Y)=\text{$\Gamma $htl}_{\mu}[\,3\,](Y,X)\nonumber\\[2mm]
&& \text{$\Gamma $htl}_{\mu}[\,3\,](R,Q)=\text{$\Gamma $htl}_{\mu}[\,5\,](Q,R) \nonumber\\
&&\text{$\Gamma $htl}_{\mu }[\,2\,](R,Q)=\text{$\Gamma $htl}_{\mu}[\,3\,](Q,R)=\text{$\Gamma $htl}_{\mu }[\,2\,](Q,R)\nonumber
\eea

\section{Explicit Vertex Expressions}\label{AppendixD}

The following results are obtained by direct calculation. They are used to rewrite Eqn. (\ref{type-1}).
\bea
&&\text{$\Gamma $htl}[\,2\,](Q,R)_{00}=\frac{\text{$\Sigma $htl}^{ret}_0(R)}{q_0}\\
&& \text{$\Gamma $htl}[\,2\,](Q,R)_{0i}=\frac{m_f^2}{r^2 q_0} \hat r_i \left(\frac{r r_0}{m_f^2} \text{$\Sigma
   $htl}^{ret}_0(R)-r\right) \nonumber\\
&&\text{$\Gamma $htl}[\,2\,](Q,R)_{ij}=\frac{m_f^2}{2 r^3 q_0} \left(\delta _{ij} \left(r r_0-\frac{r}{m_f^2}
   R^2 \text{$\Sigma $htl}^{ret}_0(R)\right)-\hat r_i \hat r_j \left(3 r r_0+\frac{r}{m_f^2} \left(r^2-3 r_0^2\right) \text{$\Sigma $htl}^{ret}_0(R)\right)\right)\nonumber\\
&&\text{$\Gamma $htl}[\,5\,](R,Q)_{00}=\frac{\text{$\Sigma $htl}^{adv}_0(R)}{q_0}\nonumber\\
&&\text{$\Gamma $htl}[\,5\,](R,Q)_{0i}=-\frac{m_f^2}{r^2 q_0}\hat r_i \left(r-\frac{r r_0}{m_f^2}
   \text{$\Sigma $htl}^{adv}_0(R)\right)\nonumber\\
&&\text{$\Gamma $htl}[\,5\,](R,Q)_{ij}=\frac{m_f^2}{2 r^3 q_0} \left(\delta _{ij} \left(r r_0-\frac{r}{m_f^2}
   R^2 \text{$\Sigma $htl}^{adv}_0(R)\right)-\hat r_i \hat r_j \left(3
   r r_0+\frac{r}{m_f^2} \left(r^2-3 r_0^2\right) \text{$\Sigma $htl}^{adv}_0(R)\right)\right)\nonumber
\eea

\section{Power Counting Arguments}\label{AppendixE}

The power counting arguments of \cite{BP} indicate that  for the real part of the self-energy, corrections to the result for the 1-loop diagram obtained by expanding to next-to-leading order in the ratio of the external momentum to the loop momentum, and contributions to the 2-loop diagrams from the region of the phase space that corresponds to both loops hard, could contribute at relative order $e$, when compared to the leading order contribution. These contributions are referred to as (1-terms) and (2-terms) in the introduction of this paper. The leading order contribution to the mass is of order $e^2 T^2/q_0 \sim eT$ (see Eqn. (\ref{leadingorder})).  Explicit calculation shows that contributions to the mass from (1-terms) and (2-terms) are not of order $e(eT) = e^2T$ but rather of order $e^3 \ln 1/e$.  A similar result was obtained by Schulz in \cite{SCH} for the gluonic plasma frequency. 

In this appendix we show why the power counting argument fails for each of the (1-terms) and (2-terms) individually, in our example. It is not known if power counting always over estimates the contributions of these terms. Throughout this appendix we use the feynman gauge. We use $K$ to denote a hard momentum, and $Q$ is the soft momentum $(q_0,\vec 0).$

Pairs of lines with the same statistics may contribute a factor $N_x(k)-N_x(k+q)\sim q/T$ and thus reduce the contribution of a diagram by one power of $e$. We show below that even without this additional power of $e$, the contributions described above do not contribute at order $e(eT)$. For the 1-loop diagram and two of the 2-loop diagrams, we make use of this knowledge in advance by considering only terms that depend on the bose-einstein distribution function. For the 2-loop bubble diagram (see Fig. 5b) it is easier to consider terms that depend on the fermi-dirac distribution function. Terms that depend on the bose-einstein distribution function contain a factor $N_B(k_0)(D_{ret}(K)-D_{adv}(K))$. This factor is even in $k_0$, is proportional to the delta function $\delta(K^2)$, and has mass dimension -2. In a power counting analysis (keeping only finite temperature contributions) we can make the replacement:
\bea
N_B(k_0)(D_{ret}(K)-D_{adv}(K))~~\rightarrow \frac{1}{k^2}\;n_b(k) \,;~~ k_0 \rightarrow n k
\eea
where the index $n$ is summed over $\pm 1$.
Similarly, for terms that contain a fermi-dirac distribution function we can write:
\bea
\label{ferm-approx}
N_F(k_0)(S_{ret}(K)-S_{adv}(K))~~\rightarrow \frac{\Ksl}{k^2}\;n_f(k) \,;~~ k_0 \rightarrow n k
\eea

We consider below all possible contributions from (1-terms) and (2-terms). The diagrams we need to consider are shown below.
\par\begin{figure}[H]
\begin{center}
\includegraphics[width=3cm]{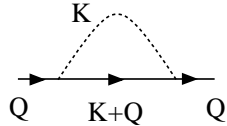}
\end{center}
\label{1loophtl}
\caption{The 1-loop htl.}
\end{figure}
\par\begin{figure}[H]
\begin{center}
\includegraphics[width=10cm]{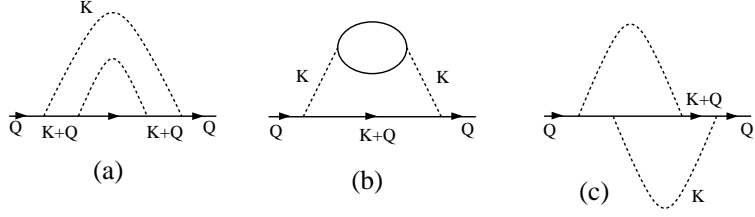}
\end{center}
\label{2loophard}
\caption{The 2-loop diagrams.}
\end{figure}
\xx (1) Corrections to the result for the 1-loop diagram obtained by expanding to next-to-leading order in the ratio of the external momentum to the loop momentum.
Consider the diagram shown in Fig. 4.
Power counting indicates that the contribution of this diagram is of order:
\bea
{\rm diagram~(1)} ~\sim~ e^2 ~\sum_{n=\pm 1}~\underbrace{~~~dk\,k^3~~~}_{{\rm phase~space}}~ \underbrace{~~~\frac{n_b(k)}{k^2}~~~}_{{\rm \gamma ~ propagator}}~ \underbrace{\frac{1}{K\cdot Q+Q^2/2}\Big|_{k_0=nk}}_{{\rm fermion ~ denominator}}~ \underbrace{{\rm Tr}(\gamma_0\gamma_\mu (\Ksl+\Qsl) \gamma^\mu)\Big|_{k_0=nk}}_{{\rm numerator}}
\eea
Doing the trace and expanding the denominator of the fermion propagator we have:
\bea
{\rm diagram~(1)} ~\sim~ e^2~\sum_{n=\pm 1}~ dk ~k ~n_b(k)~\frac{1}{q_0}\Big(1+\frac{q_0}{k_0}\Big) ~\Big(1-\frac{q_0^2}{2k_0 q_0}+\cdots \Big)\Big|_{k_0=nk}
\eea
The leading order contribution is:
\bea
{\rm diagram~(1)}_{lo} ~\sim~ e^2 ~ dk ~k ~n_b(k)~\frac{1}{q_0}~\sim~ e^2\frac{T^2}{q_0}
\eea
which agrees with (\ref{leadingorder}). 
The next term in the expansion gives zero when summed over $n$ and thus the term of relative order $e$ that is predicted by the power counting arguments of \cite{BP} identically cancels.
\bea
{\rm diagram~(1)}_{nlo}~\sim~ e^2 \sum_{n=\pm 1}~ dk~ n ~n_b(k) ~\rightarrow e^2T \cdot (0)
\eea
The largest sub-leading term is of order:
\bea
{\rm diagram~(1)}_{nnlo}~~\sim~ e^2 q_0~\frac{dk}{k} ~n_b(k) ~\rightarrow~~ e^2 q_0~dk~  n_b(k)~k ~{\rm Prin}\frac{1}{k^2-q_0^2} \sim e^3 T \ln (1/e)
\eea
where the arrow indicates the expression that would be obtained if the expansion was done carefully enough to preserve the term that regulates the apparent infra-red divergence in the log.\\

\xx (2) Contributions from the double-rainbow diagram with both loop momenta hard.

\xx Consider the diagram in Fig. 5a. Since both loop momenta are hard, we can replace the inner loop by the leading order piece of the 1-loop self-energy. The leading order contribution is:
\bea
{\rm diagram~(2)} ~\sim~ e^2 ~\sum_{n=\pm 1}~\underbrace{~~~dk\,k^3~~~}_{{\rm phase~space}}~ \underbrace{~~~\frac{n_b(k)}{k^2}~~~}_{{\rm \gamma ~ propagator}}~ \underbrace{\frac{1}{(K\cdot Q+Q^2/2)^2}\Big|_{k_0=nk}}_{{\rm fermion ~ denominators}}~\underbrace{{\rm Tr}(\gamma_0\gamma_\mu\Ksl \gamma_\tau \Ksl \gamma^\mu)\Sigma{\rm htl}^\tau(K)\Big|_{k_0=n k}}_{numerator}
\eea
If we collect the powers of $e$ it appears as if we have: an explicit factor of $e^2$ in front; a factor $e^2$ contained in the htl self-energy; and a factor of $1/e^2$ from the 2 powers of $1/q_0$ introduced by the 2 fermion propagators. By simple dimensional analysis this should produce a contribution $e(eT)$ as predicted in \cite{BP}. However, if we look closely at the numerator of this expression we see that, as in case (1) above, this leading order term cancels identically.
If we do the trace we obtain two kinds of terms:
\bea
&& (i)~~~~k_0 (K_\mu \Sigma{\rm htl}^\mu(K))\Big|_{k_0=n k} \sim n~ k~ e^2 T^2 \rightarrow 0 \\
&& (ii)~~~~K^2 \Sigma{\rm htl}_0(K)|_{k_0=n k} \rightarrow 0 \nonumber
\eea
The first term is zero when we sum over $n$ and the second term is zero when we substitute in for $k_0$.\\

\xx (3) Contributions from the bubble diagram with both loop momenta hard.

\xx Consider the diagram in Fig. 5b. Since both loop momenta are hard, we can replace the bubble by the leading order piece of the 1-loop photon polarization tensor. In this case it is easier to consider the terms that depend on the fermi-dirac distribution function. 
Using (\ref{ferm-approx}) the leading order contribution is:
\bea
{\rm diagram~(3)} ~\sim~ e^2 ~\sum_{n=\pm 1}~\underbrace{~~~dk\,k^3~~~}_{{\rm phase~space}} \underbrace{~~\frac{n_f(k)}{k^2}~~}_{{\rm fermion ~denominator}}\underbrace{\frac{1}{(K\cdot Q+Q^2/2)^2}\Big|_{k_0=nk}}_{{\rm \gamma~ propagators}} ~\underbrace{{\rm Tr}(\gamma_0\gamma_\mu\Ksl \gamma_\nu)g^{\mu\tau}\Pi_{\tau\lambda}g^{\lambda\nu}\Big|_{k_0=nk}}_{numerator}
\eea
If we collect the powers of $e$ it appears as if we have: an explicit factor of $e^2$ in front; a factor $e^2$ contained in the htl polarization tensor; and a factor of $1/e^2$ from the 2 powers of $1/q_0$ introduced by the 2 photon propagators. By dimensional analysis this should produce a contribution $e(eT)$ as predicted in \cite{BP}. However, if we look closely at the numerator of this expression we see that, as in cases (1) and (2) above, this leading order term cancels identically. The htl polarization tensor contains three different structures: $g_{\tau\lambda}$, $K_\tau K_\lambda/K^2$ and $k_0 \{g_{\tau 0} K_\lambda,g_{\lambda 0} K_\tau\}$. Doing the trace and contracting with each of these tensors we find that the first two produce a term proportional to $k_0$ (which gives zero when summed over the index $n$), and the third is proportional to $K^2/k^2$ (which gives zero when we substitute in for $k_0$).\\

\xx (4) Contributions to the crossed-rainbow diagram with both loop momenta hard.

\xx Consider the diagram in Fig. 5c. Since both loop momenta are hard, we can replace the left hand loop by the leading order piece of the htl vertex. The leading order contribution is:
\bea
{\rm diagram~(4)} ~\sim~ e^2 ~\sum_{n=\pm 1}~\underbrace{~~~dk\,k^3~~~}_{{\rm phase~space}}~ \underbrace{~~~\frac{n_b(k)}{k^2}~~~}_{{\rm \gamma ~ propagator}}~ \underbrace{\frac{1}{(K\cdot Q+Q^2/2)}\Big|_{k_0=nk}}_{{\rm fermion ~ denominator}}~\underbrace{{\rm Tr}(\gamma_0\gamma_\tau \Ksl \gamma_\mu)\Gamma{\rm htl}^{\tau\mu}(K)\Big|_{k_0=n k}}_{numerator}
\eea
There are not enough powers of $1/q_0$ to produce a contribution of order $e(eT)$.\\

The conclusion is that, for the case of a static fermion, the power counting argument of \cite{BP} over estimates the contribution of: (1) the subleading corrections to the 1-loop htl result and; (2) each of the 2-loop diagrams with both loop momenta hard. In every case, the nlo terms predicted by \cite{BP} are present, but they cancel exactly at the relative $e$ order that is being considered.\\

\Large 

\xx {\bf Acknowledgments}\\

\normalsize

I am very grateful to Doug Pickering for help with the numerical calculations. 
I thank Emil Mottola for discussions. 
This research was supported by the Natural Sciences and
Engineering Research Council of Canada.

\end{document}